# ARTIFICIAL INTELLIGENCE TECHNOLOGY IN ONCOLOGY:
# A NEW TECHNOLOGICAL PARADIGM

Mario Coccia

National Research Council of Italy & Yale University

Yale University School of Medicine, Global Oncology, Yale Comprehensive Cancer Center

310 Cedar Street, Lauder Hall, Suite 118, New Haven, CT 06510

E-mail: mario.coccia@cnr.it

**Abstract.** Artificial Intelligence (AI) technology is based on theory and development of computer systems able to perform tasks that normally require human intelligence. In this context, deep learning is a family of computational methods that allow an algorithm to program itself by learning from a large set of examples that demonstrate the desired behavior. Application of these methods to medical imaging can assist pathologists in the detection of cancer subtype, gene mutations and/or metastases for applying appropriate therapies. The purpose of this study is to show the emerging application of AI in medical imaging to detect lung and breast cancer. Moreover, this study shows the comparative evolutionary pathways of this emerging technology for three critical cancers: lung, breast and thyroid. A main finding of this study is the recognition that, since the late 1990, the sharp increase of technological trajectories of AI technology applied in cancer imaging seems to be driven by high rates of mortality of some types of cancer (e.g., lung and breast) in order to find new techniques for a more accurate detection, characterization and monitoring as well as to apply efficiently anticancer therapies that increase the progression-free survival of patients: the so-called mortality-driven AI technological trajectories. Results also suggest that this new technology can generate a technological paradigm shift for diagnostic assessment of any cancer type. However, application of these methods to medical imaging requires further assessment and validation to assist pathologists to increase the efficiency of their workflow in both routine tasks and critical cases of diagnostics.

Acknowledgement. I also gratefully acknowledge financial support from National Research Council of Italy–Direzione Generale Relazioni Internazionali for funding this research project developed at Yale University in 2019 (research project ammcnt-cnr n. 62489, 26 September 2018). Usual disclaimer applies.

**Introduction**

Artificial Intelligence (AI) is the research field that studies and develops technological systems that can solve complex tasks in ways that would traditionally need human intelligence (Iafrate, 2018). This new technology, originated from interaction between engineering, computer science and other applied sciences, is developing main fields of applied research with technology transfer in robotics, natural language processing, machine learning, computer vision, etc. The artificial intelligence can be the source of a self-propagating development of technology that acquiring new functionality from digital industry and information and communication technology, it can generate a vast industrial and social change in economic systems (cf., Geuna et al., 2017). In this domain of AI, Machine Learning (ML) has made a tremendous progress over the last decades. Many scholars believe that ML is capable of making progress on any real-world problem. In the research field of machine learning, deep learning is a family of computational methods that allow an algorithm to program itself by learning from a large set of examples that demonstrate the desired behavior, removing the need to specify rules explicitly (cf., Goodfellow et al., 2018).

In this context, the paper here shows the first applications of AI technology to cancer imaging that can generate a technological paradigm shift in oncology for improving diagnostics and accelerating treatments of these diseases (cf., Kantarjian and Yu, 2015; Li et al., 2019). This study focused on case studies of new deep learning algorithms for detection of lung and breast cancer that are two of the most important diseases in society. These case studies can show the high potential of path-breaking applications of AI in medicine that can generate an industrial and social change in clinics. Moreover, this study analyses the comparative evolutionary pathways of this emerging technology for three critical cancers (lung, breast and thyroid) to explain possible drivers. A discussion shows socioeconomic barriers to the diffusion of this new technology in medicine that can be generalized for explaining the dynamics of this new technology in medicine and next applications in other sectors of economic system.

**Background**

Artificial Intelligence (AI) is the research field that studies and develops technological systems that can solve complex tasks in ways that would traditionally need human intelligence. The artificial intelligence can generate a vast industrial and social change in economic systems (cf., Geuna et al., 2017).

In this domain of AI, Machine learning (ML) is the scientific study of algorithms and statistical models that computer systems use to effectively perform a specific task without using explicit instructions, relying on patterns and inference instead (Fig. 1). Machine learning algorithms build a mathematical model based on sample data, known as "training data", in order to make predictions or decisions without being explicitly programmed to perform the task.

Deep learning is part of a broader family of machine learning methods based on artificial neural networks (Madabhushi et al., 2016). Deep learning architectures such as deep neural networks, deep belief networks, recurrent neural networks and convolutional neural networks have been applied to fields including computer vision, speech recognition, natural language processing, audio recognition, social network filtering, machine translation, bioinformatics, drug design, medical image analysis, histopathological diagnosis, material inspection and board game programs, where they have produced results comparable to and in some cases superior to human experts (Litjens et al., 2016).

This new technology, in particular, deep learning algorithms can be applied for improving diagnostic accuracy and efficiency of detection of cancer.

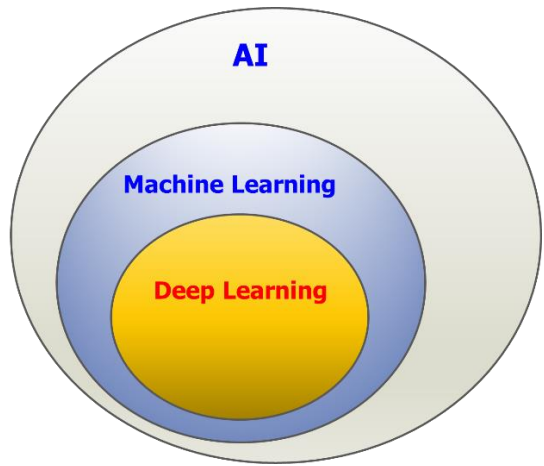

**Figure 1.** Artificial Intelligence and it main research fields

Cancer is an organism which lives off a host organ, growing by bio-genetic-molecular mechanism. 9.6 million people died of cancer in 2018 – more than from HIV/AIDS, malaria and tuberculosis combined. The incidence of cancer is estimated to double by 2035, with most of these cases expected to occur in low-to-middle income countries (LMICs). 60% of cancer cases occur in LMICs, and 75% cancer deaths occur in these countries (Prager, 2018). In many LMICs, breast cancer remains the leading malignancy affecting women and the leading cause of cancer-related deaths (Bray et al., 2018). Table 1 shows that lung and breast cancer have the highest mortality rate worldwide (cf. also Parkin *et al.*, 2005; Coccia, 2015).

Table 1. Incidence and mortality of big 8 cancers in 2018, worldwide, both sexes, all ages

| Cancer | Incidence ASR (W)* | Mortality ASR (W)* |
|---|---|---|
| **Lung** | **22.5** | **18.6** |
| **Breast** | **46.3** | **13.0** |
| Colorectum | 19.7 | 8.9 |
| Prostate | 29.3 | 7.6 |
| Stomach | 11.1 | 8.2 |
| Pancreas | 4.8 | 4.4 |
| Ovary | 7.8 | 3.9 |
| Liver | 9.3 | 8.5 |

*Note*: Age-Standardized Rate-ASR (W): A rate is the number of new cases or deaths per 100,000 persons per year. An age-standardized rate is the rate that a population would have if it had a standard age structure.

*Source*. World Health Organization, International Agency for Research on Cancer (2019).

The R&D in oncology is supporting the convergence of different research fields, such as genetics[1], genomics[2], nanotechnology, nanomedicine, computer sciences, etc., that are generating new technological pathways for diagnostics and therapeutics (Coccia, 2014, 2015a, 2016, 2018). Clinical challenges are in the accurate detection and characterization of cancers. In particular, artificial intelligence (AI), generated by converging technologies in applied sciences, can support the qualitative interpretation of cancer imaging by expert clinicians, such as volumetric delineation of tumors as well as evaluation of the impact of disease and treatment on adjacent organs. In general, AI may automate processes in the initial interpretation of images and shift the clinical workflow on high-level decisions to efficiently apply anti-cancer treatments for patients.

---
[1] Genetics studies the molecular structure and function of genes in the context of a cell or organism.
[2] Genomics is a discipline in genetics that studies the genomes of organisms. In particular, it determines the entire DNA sequence of organisms and fine-scale genetic mapping efforts.

**The goal of this investigation**

The purpose of this study is to show the potential of Artificial Intelligence Technology in cancer imaging to distinguish between cancer, metastases or normal lung tissue. This study shows two case study of clinical challenges and applications of AI in lung and breast cancer. Moreover, this study endeavors to analyze evolutionary pathways of this AI for three critical cancers (lung, breast and thyroid) to explain possible drivers.

**Importance**

Importance of the paper here is to show that this emerging technology applied in medicine shows increasingly concerted efforts in pushing AI technology to clinical use and to impact future directions in cancer care, although most studies evaluating AI applications in oncology to medical imaging requires further assessment and validation for reproducibility and generalizability of results.

**Material and Methods: Study design and setting**

In order to show the potential of deep learning algorithms in cancer imaging, this study describes two case studies of AI technology applied for detection and characterization of lung cancer and lymph node metastases in women breast cancer.

Moreover, to show the evolutionary pathways of AI technology and deep learning in the research field of cancer, this study considers data from ScienceDirect (2019) using the tool Advanced Search to find articles that have in title, abstract or keyword the following terms:

- Artificial intelligence and lung cancer
- Artificial intelligence and breast cancer
- Artificial intelligence and thyroid cancer
- Deep learning and lung cancer

- Deep learning and breast cancer

- Deep learning and thyroid cancer

The evolution of AI technology in cancer settings is measured with the number of articles from 1996 to 2018 for a comparative analysis of the trends of three different types of cancer (lung, breast and thyroid). The visual representation of trends is also used to detect the most appropriate model to estimate the relation of number of articles over time and assess the evolutionary growth of this new technology between different research fields of oncology (cf., Coccia, 2005, 2018a, 2019).

The relation between number of articles per years of AI applied in cancer imaging of different cancers is estimated with regression analysis (Ordinary Least Squares Method) using an exponential model. Outputs of statistical analysis are: regression coefficients, adjusted $R^2$, standard error of the estimate and analysis-of-variance table.

In order to assess the evolutionary growth of AI technology applied to specific types of cancer (lung, breast and thyroid cancer) the following exponential equation is applied over 1996-2018 period:

$P_{t(2018)} = P_{0(1996)} \, e^{rt}$

P= number of articles with a combination of keywords

$\frac{P_t}{P_0} = e^{rt}$

$Log\left(\frac{P_t}{P_0}\right) = r \cdot t$

$r = \frac{Log\left(\frac{P_t}{P_0}\right)}{t}$

A comparative analysis of results can suggest the direction and intensity of acceleration of technological trajectories of AI technology in different cancer settings and possible drivers.

**Results: case studies and evolution of AI technology in cancer imaging**

- *Artificial Intelligence and deep learning for detecting the typology of lung cancer: a revolution in medicine*

Lung cancer is one of the main diseases in several countries and a leading cause of cancer death – both sexes – worldwide. Lung cancer is linked to several risk factors in society, such as smoking, passive smoking, air pollution, etc. Lung cancer can be either small cell lung cancer or no small cell lung cancer (NSCLC), with the latter representing about 80% of the cases. The mortality of lung cancer is very high and the five-year survival rate of patients is about 2–10% (Coccia, 2014, 2012). The detection of the type of lung cancer (e.g., adenocarcinoma, squamous cell carcinoma, etc.), mutations (e.g., EGFR: epidermal growth factor receptor; ALK: anaplastic lymphoma receptor tyrosine kinase rearrangements) and sequential mutations (T790) is a critical process in diagnostics to select the appropriate therapy that can be conventional chemotherapies, target therapies such as gefitinib, erlotinib, etc., or new immune therapies, e.g., nivolumab, atezolizumab (cf., Coccia, 2014). This critical process affects the survival of patients (cf., Fig. 2).

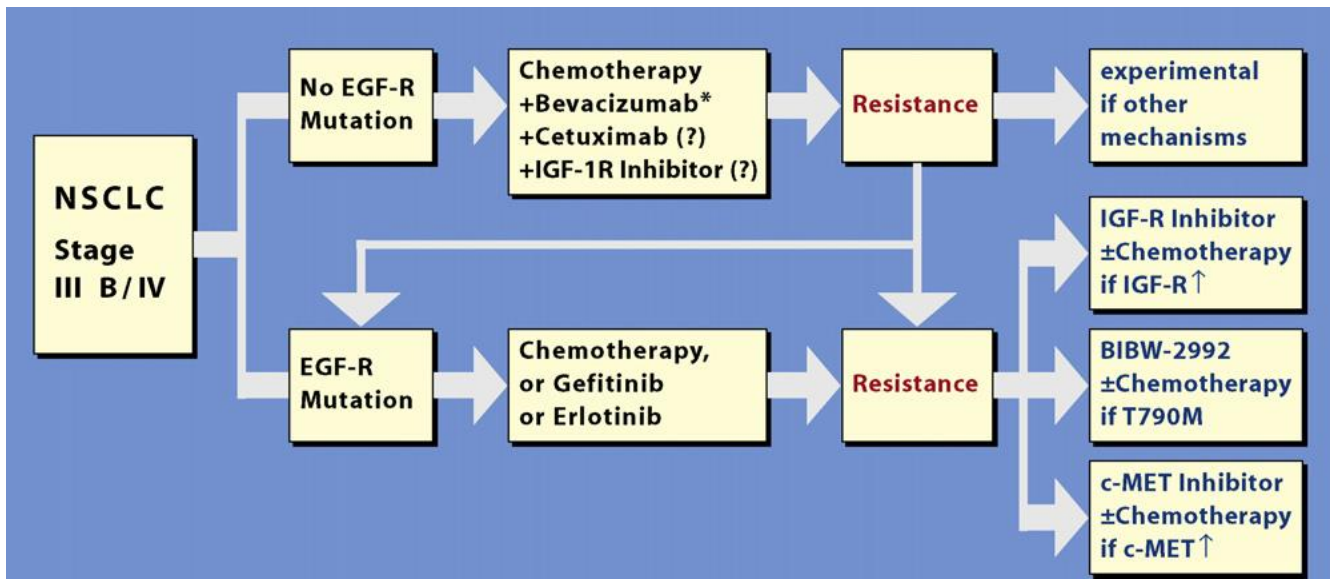

Figure 2. Roadmap for Non-small lung cancer treatments based on EGF-R blocking agents. *Source*: Dempke et al. (2010), Lung Cancer, 67, p. 265

Classification and mutation of lung cancer type is a critical diagnostic process because the available treatment options, including conventional chemotherapy and, more recently, targeted therapies, differ for lung adenocarcinoma and lung squamous cell carcinomas.

Current technology for an accurate diagnosis and selection of treatment options is *based on molecular biomarkers applied on lung biopsies* and/or *blood testing*: this approach can diagnose lung cancer type and stage. This current technology has some negative effects. On the one side, patients have to receive an invasive surgical intervention for tissue biopsies or to do liquid biopsy (e.g., blood) and wait a certain time (about a month) before to have the diagnosis of cancer type and correct treatments, and the time in the presence of cancer is a critical variable for survival and, when possible, for healing. On the other side, hospitals have a high cost with these approaches that require appropriate equipment and specialized personnel, which in some regions are scarce or overload of work because of shrinking public lab and hospital budgets.

New technology is based on artificial intelligence technology and some studies show that deep learning can be used for the classification of breast, bladder and lung tumors (Khosravi et al., 2018). In particular, the development of new, inexpensive and more powerful artificial intelligence technologies has made possible the training of larger and more complex neural networks. This has resulted in the design of several deep convolutional neural networks (CNNs) that are capable of accomplishing complex visual recognition tasks. Coudray et al. (2018) show that medicine can benefit from deep learning technology by convolutional neural networks (CNNs) that not only outperforms other methods, but also achieves accuracies to classify the type of cancer that are comparable to pathologists for selecting the appropriate therapy. These models maintain their performance when tested on independent datasets of frozen tissues as well as on images obtained from biopsies. CNNs have been applied with regard to classification of lung patterns on computerized tomography (CT) scans, with positive results. This new technology is due to Google Corporation that in 2014 won the ImageNet Large-Scale Visual Recognition Challenge. The ImageNet project is a large visual database designed for use in visual object recognition software research. More than 14 million images have been hand-annotated by the project to indicate what objects are pictured and in at least one million of the images, bounding boxes are also provided. ImageNet, developing the GoogleNet architecture, increased the robustness to translation and nonlinear learning abilities by using microarchitecture units called inception. Each inception unit includes several nonlinear convolution modules at various resolutions (cf., Russakovsky et al., 2015). Inception architecture is particularly useful for processing the data in multiple resolutions, a feature that makes this architecture suitable for medicine and in particular for pathology tasks. This complex neural network has already been successfully adapted to specific types of disease classifications like, skin cancers and diabetic retinopathy detection (Esteva et al., 2016; Gulshan et al., 2016). Coudray et al. (2018) have developed a deep-learning model for the automatic analysis of tumor slides using publicly available whole-slide images available in The Cancer Genome Atlas (TCGA). These scholars trained inception v3 to recognize tumor in lung versus normal tissue. After that, they tested the performance of these methods on the more challenging task of distinguishing lung adenocarcinoma and lung squamous cell carcinoma. They also evaluated the deep-learning model by training and testing the network on a direct three-way classification into the three types of images (normal tissue, lung adenocarcinoma, lung squamous cell carcinoma). Results of this approach are compared to the evaluation of three pathologists (two thoracic pathologists and one anatomic pathologist) that independently classify the whole-slide images in the test set by visual inspection alone, independently of the classification provided by TCGA. Overall, the performance of deep learning models was comparable to that of each pathologist. Hence, new artificial intelligence technology, based on deep-learning models, can assist pathologists in the detection of cancer subtype or gene mutations in any cancer type with a save of time and costs; moreover, poor regions, with these AI technologies, can also have high benefits by sending the digital images to labs of other countries, generating a reduction of current gap in healthcare between countries.

⬜ *Deep learning for detection of metastases in women with breast cancer*

Breast cancer is the most frequent cancer worldwide among women (see tab. 1). Studies based on advanced countries show that the incidence of breast cancer has been increasing. It has been postulated that breast cancer incidence tends to be higher in more developed countries due to delayed childbearing, a higher use of hormone replacement therapy, a higher rate of screening, and improved tumor registries. Some studies have also argued that higher income countries may have higher fat diets and an increased rate of obesity, both correlated with higher breast cancer incidence rates. In general, scholars note that many Western populations have a higher incidence rate of breast cancer than regions in Africa and Asia (Coccia, 2013).

Breast biopsies, like all types of cancer, are used to diagnose breast cancer type, stage and mutations. An accurate breast cancer staging is an essential task performed by pathologists to inform clinical management. For instance, evaluation of the extent of cancer spread by histopathological analysis of sentinel axillary lymph nodes is an important part of breast cancer staging. The sensitivity of sentinel axillary lymph nodes evaluation by pathologists, however, is sub-optimal because some studies show that pathology review by experts changed the nodal status in 24% of patients (Ehteshami Bejnordi et al., 2017). This effect is due to the difficulty of the process of decision making of human behavior. Moreover, sentinel axillary lymph nodes evaluation is time-consuming and has a certain cost. In short, the high cost of exams and a period of more than one month to have breast cancer type, stage, etc. to decide the appropriate therapy, it can affect results of therapeutic treatments and survival of patients.

The application of Artificial Intelligence (AI) technology with deep learning algorithms to whole-slide pathology images can potentially improve diagnostic accuracy of breast cancer and metastases (Lamy et al., 2019). Some scholars have assessed the performance of automated deep learning algorithms at detecting metastases in tissue sections of lymph nodes of women with breast cancer and compared results with pathologists' diagnoses. In the setting of a challenge competition, some deep learning algorithms achieved better diagnostic performance than a panel of 11 pathologists participating in a simulation exercise designed to mimic routine pathology workflow; in short, algorithm performance was comparable with an expert pathologist interpreting whole-slide images without time constraints (Ehteshami Bejnordi et al., 2017). Hence, this experiment has shown that deep learning algorithms could identify metastases in sentinel axillary lymph nodes slides with 100% sensitivity, whereas 40% of the slides without metastases could be identified as such. This approach can significantly reduce the workload of pathologists and improve the management decisions on whether or not to administer a therapy, perform a surgical intervention, etc. Overall, then, this interesting result shows the potential of AI applied in cancer imaging for detection of lymph node metastases in women with breast cancer to improve processes of diagnosis and therapy of this and other critical cancers (cf., Ehteshami Bejnordi et al., 2017).

- *Statistical analyses of evolutionary trajectories of AI technology in cancer imaging (lung, breast and thyroid).*

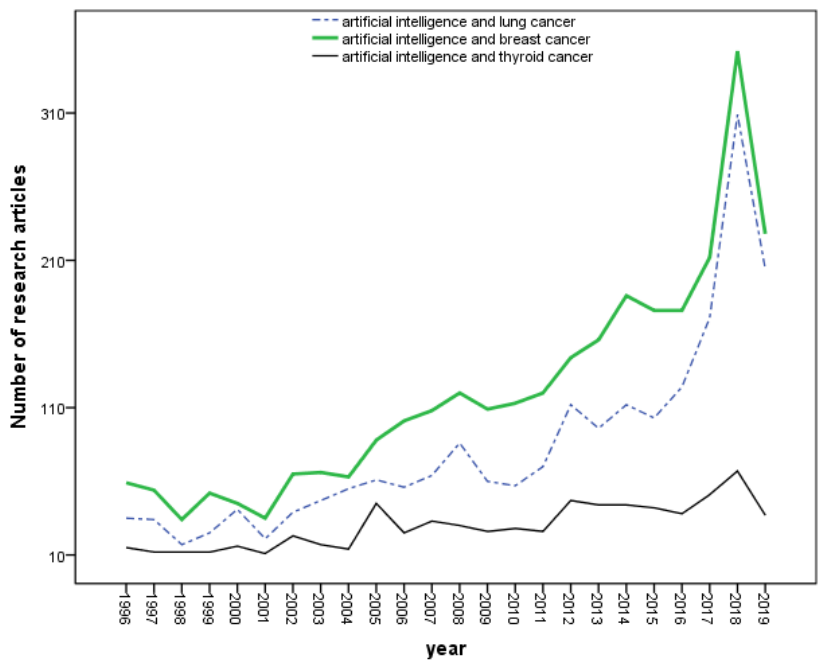

Figure 3. Evolutionary trends of number of article about AI technology per different types of cancer (lung, breast, and thyroid) over 1996-2018 period.

Table 2. Estimated relationships of AI in articles per different types of cancer on time (Exponential model, 1996-2018 period)

|  | Lung cancer | Breast cancer | Thyroid cancer |
|---|---|---|---|
| *Constant $\lambda_0$* | 1.000E-013 | 1.000E-013 | 1.000E-013 |
| *(St. Err.)* | (0.00) | (0.00) | (0.00) |
|  |  |  |  |
| *Coefficient $\lambda_1$* | 0.093*** | 0.083*** | 0.067*** |
| *(St. Err.)* | (0.008) | (0.006) | (0.008) |
|  |  |  |  |
| *F* | 127.46 | 186.63 | 72.34 |
| *Sig.* | 0.001 | 0.001 | 0.001 |
|  |  |  |  |
| *$R^2$ adj.* | 0.85 | 0.89 | 0.76 |
| *(St. Err. of the Estimate)* | (0.28) | (0.21) | (0.28) |

*Note*: Dependent variable: Number of articles about the terms Artificial intelligence per different types of cancer; explanatory variable is time; ***= *p-value*< .001

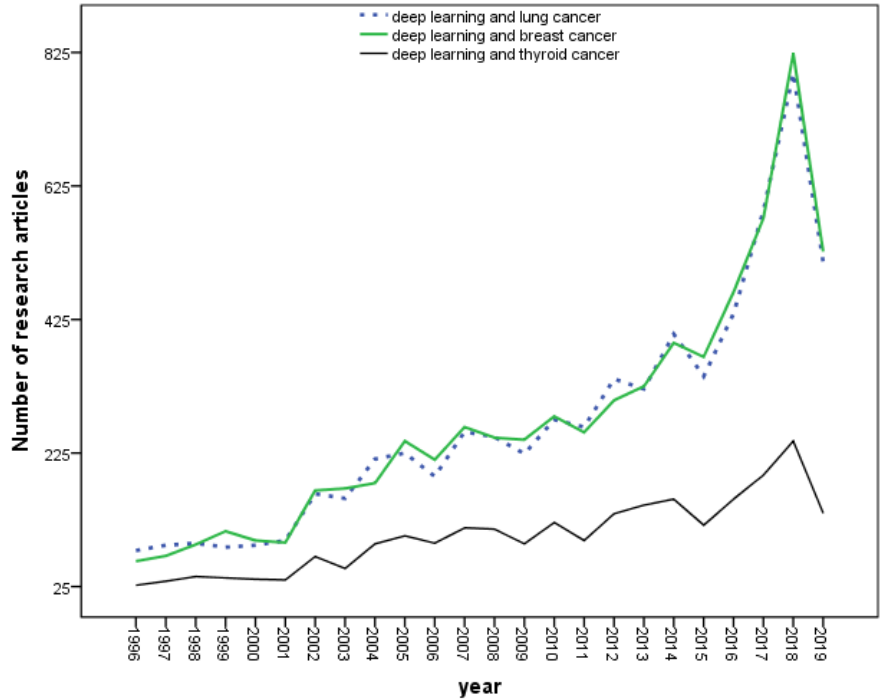

Figure 4. Evolutionary trends of number of article about deep learning algorithms per different types of cancer (lung, breast, and thyroid) over 1996-2018 period.

Table 3. Estimated relationships of deep learning articles per different types of cancer on time (Exponential model, 1996-2018 period)

|  | *Lung cancer* | *Breast cancer* | *Thyroid cancer* |
|---|---|---|---|
| *Constant $\lambda_0$* | 1.000E-013 | 1.000E-013 | 1.000E-013 |
| *(St. Err.)* | (0.00) | (0.00) | (0.00) |
| *Coefficient $\lambda_1$* | 0.091*** | 0.094*** | 0.082*** |
| *(St. Err.)* | (0.005) | (0.005) | (0.007) |
| *F* | 302.55 | 317.25 | 138.49 |
| *Sig.* | 0.001 | 0.001 | 0.001 |
| *$R^2$ adj.* | 0.92 | 0.93 | 0.86 |
| *(St. Err. of the Estimate)* | (0.18) | (0.18) | (0.23) |

*Note*: Dependent variable: Number of articles about the terms Deep Learning per different types of cancer; explanatory variable is time; ***= *p*-value< .001

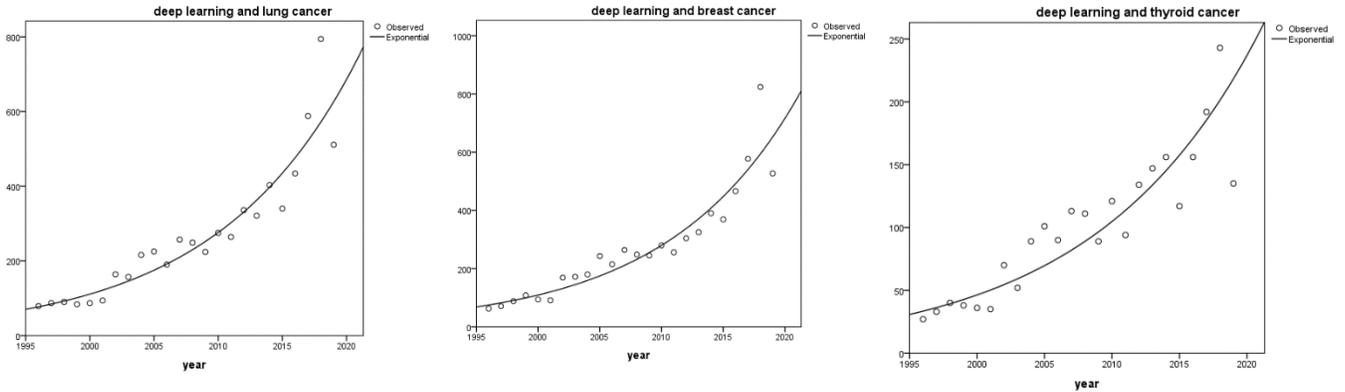

Figure 5. Exponential curves of the number of articles about deep learning per different types of cancer (lung, breast, thyroid) on time, 1996-2018 period

Table 4 – Rates (per cent) of evolutionary pathways (based on exponential growth) of artificial intelligence and deep learning in lung, breast and thyroid cancer from 1996 to 2018

| Rates of growth | Artificial Intelligence and lung cancer | Artificial Intelligence and breast cancer | Artificial Intelligence and thyroid cancer | Deep learning and lung cancer | Deep learning and breast cancer | Deep learning and thyroid cancer |
|---|---|---|---|---|---|---|
| $r$ exponential | **9.899** | **8.119** | 6.803 | **10.489** | **11.687** | 9.987 |

Figures 3 - 4 show that the evolution of AI in oncology has exponential patterns of growth started in 1990s and ongoing, with an acceleration mainly for lung cancer (cf., Fig. 5). Statistical analysis based on regression analysis shows that $R^2$ adjusted of models indicates that more than 85% of variation in production of articles in AI per different types of cancer can be attributed (linearly) to the time as predictor (Tabb. 2-4). In particular, estimated relationships and exponential rates of growth suggest that AI in cancer imaging of breast and lung have a faster growth than thyroid cancer.

A main finding of this study is the recognition that, since the late 1990s, the sharp increase of technological trajectories of AI technology applied in cancer imaging seems to be driven by high rates of mortality of some types of cancers (e.g., lung and breast) in order to find techniques for a more accurate detection, characterization and monitoring to apply efficiently anticancer therapies that increase the progression-free survival of patients: the so-called mortality-driven technological trajectories in AI (cf., Coccia and Wang, 2014; Coccia, 2016). These ground-breaking technological trajectories of AI are paving new directions and clinical challenges, creating the foundations for a revolution in medicine that may lead to more effective diagnostic equipment in clinical settings in the not-too-distant future for a paradigm shift in clinical practice of pathology.

**Limitations**

The results and arguments of this study are of course tentative. In fact, the phenomenon is complex and analyses here are not sufficient to explain the pros and cons of the patterns of AI in cancer imaging, since we know that other things are often not equal over time and space. This study may form a ground work for development of more sophisticated studies and theoretical frameworks. Future efforts in this research field should provide more statistical evidence to substantiate the theoretical framework here. To reiterate, the study here is exploratory in nature and there is need for much more detailed research to shed further theoretical and empirical light on technological trajectories supporting the clinical challenges and applications of AI technology in cancer imaging. This analysis shows general pathways of the evolution of AI in oncology, however the evolution of AI technology in medicine is a non-trivial exercise because it depends on the behavior of other technologies and manifold factors of socioeconomic systems. In this context, Wright (1997, p. 1562) properly claims that: "In the world of technological change, bounded rationality is the rule."

**Discussion and conclusion**

Bi Wenya et al. (2019) argue that cancer offers a unique context for medical decisions based on variegated forms of disease, individual condition of patients, their ability to receive treatment, and their responses to treatment (cf., Rosenberg, 1969; Arthur, 2009; McNerney et al., 2011). Clinical challenges are in the accurate detection, stage, type and sub-type of cancers. Radiographic evaluation of cancer most commonly relies upon visual analysis of imaging, the interpretations of which may be augmented by advanced computational analyses. The new technology of artificial intelligence (AI) can improve the interpretation of cancer imaging by pathologists to detect volumetric delineation of tumors over time, mutations, and diffusion on other organs. AI may support the clinical process of interpretation of images to shift the clinical workflow of pathologists towards management decisions on whether or not to administer an intervention, and/or anticancer drugs, etc.

In medicine, radiology, having converted to digital images about 30 years ago, is well-positioned to deploy AI for diagnostics. Several studies have shown considerable opportunity to support radiologists in evaluating a variety of scan types, including mammography for breast lesions, computed tomographic scans for pulmonary nodules and infections, and magnetic resonance images for brain tumors including the molecular classification of brain tumors. With conversion to digital images, radiology reduced costs eliminating film, chemicals, developers, and storage of the films. Moreover, radiology departments also solved problems related to loss of films and transport of films to where they are needed, for example, in operating rooms, emergency departments, and intensive care units. In short, digital images using computers have improved the quality, safety, and efficiency of radiologists in the management of disease.

Pathology, by contrast to radiology, has been late to adopt digital imaging and computer-assisted diagnostic technologies. The emergence of AI in health care, the reduced costs of digital data, and the availability of usable digital images are now in alignment for digital pathology to succeed.

However, there are some barriers to the technological diffusion of AI in some medical fields. A major unresolved issue is how AI will be implemented in routine clinical practice. Numerous intertwined problems will have to be addressed to overcome several significant obstacles for the diffusion of AI in pathology and other branches of medical science.

*Firstly*, digital pathology based on AI may likely be costly at the initial phase of diffusion of this new technology owing to additional workflow, including personnel, AI must demonstrate improved efficiency, quality, and safety. These issues are the first and immediate barriers to a broad application of AI in pathology and clinical practice. However, benefits will inevitably come from the use of AI with digital images and multiple data sets, for example, integrating genomic data and radiologic images to further enhance the value of combined data utilization for the overall health care system. Recent examples with skin lesions, diabetic retinopathy, and radiology detection have highlighted the potential that AI provides to aid clinicians to improve quality, safety and diagnosis (Gulshan et al., 2016; Esteva. et al. 2016)

*Secondly*, the cost of new technological devices is an economic barrier. This cost is associated to the efficacy and efficiency of AI in pathology that have to be clearly demonstrated to government and third-party payers (e.g., insurance companies) developing reimbursement strategies. Although some reimbursement codes exist for computational analyses in the USA, they are not widely used and often are rejected. With US national health care reimbursement trends moving to quality and safety metrics, the recognition of AI technology as part of reimbursement strategies that reward value-based care, rather than fee for service value-based care, would provide important incentives to develop and implement validated algorithms of AI in clinical practice.

Thirdly, the critical factor for AI technology is human capital and education that will be the greatest challenge and will require the longest period to address. AI and other computational methods must be integrated into all training programs of the schools of medicine worldwide. Future generations of pathologists and physicians must have technical skills comfortable to use digital images and other data in combination for accurate diagnosis.

Overall, then, the promise of AI in health care is the delivery of improved quality and safety of care with the great potential of reducing the inequality in the access to healthcare sector between people within poor and rich countries. Put otherwise, AI technology in medicine can increase the democratization of the health care system, and it can potentially provide value in many economic sectors generating benefits for an industrial and corporate change based on this new technological paradigm (cf., Coccia, 2010, 2014a). For instance, radiologists can now read imaging studies from anywhere in the world at their home institution/office, bringing expert care to parts of the world that previously had limited expertise. Pathology has the opportunity to do the same with digital imaging and AI combined with ICTs can permit rapid and accurate local care, also in developing countries where there is a high burden of diseases and low medical equipment.

To conclude, deep learning and in general AI have the opportunity to assist pathologists and physicians by improving the efficiency of their work, standardizing quality, and providing better prognostication. Although workflow of pathologists and other physicians is likely to change with AI technology, the contributions of pathologists, and in general of humans, to patient care will continue to be critically important to treat diseases. Finally, most studies evaluating AI in oncology and other branches of medicinal sciences to date have not been vigorously validated for reproducibility and generalizability of results. Further research is necessary in applied research to determine the feasibility of applying AI in the clinical setting and to determine whether use of the AI could lead to improved care and outcomes compared with current technologies and medical assessments. Overall, then, the technological change based on technological paradigm shift of AI in medicine seems to be started to push regional and national economies in the next future.